\documentclass[twocolumn,prb, aps,floatfix,nobibnotes]{revtex4-2}
\usepackage{amsmath}
\usepackage[utf8]{inputenc}
\usepackage{graphicx}
\usepackage{braket}
\usepackage[colorlinks=true,citecolor=blue]{hyperref}

\newcommand{\ex}{{\hat{\boldsymbol{x}}}}
\newcommand{\ey}{{\hat{\boldsymbol{y}}}}

\newcommand{\en}{{\hat{\boldsymbol{n}}}}
\newcommand{\ha}{{\hat{a}}}
\newcommand{\hb}{{\hat{b}}}
\newcommand{\hc}{{\hat{c}}}
\newcommand{\hn}{{\hat{n}}}
\newcommand{\hal}{{\hat{\alpha}}}
\newcommand{\hbe}{{\hat{\beta}}}
\newcommand{\hnu}{{\hat{\nu}}}
\newcommand{\hmu}{{\hat{\mu}}}
\newcommand{\hrho}{{\hat{\rho}}}
\newcommand{\cH}{{\mathcal{H}}}
\newcommand{\ctH}{\tilde{{\mathcal{H}}}}
\newcommand{\cL}{{\mathcal{L}}}
\newcommand{\bH}{{\boldsymbol{H}}}
\newcommand{\bh}{{\boldsymbol{h}}}
\newcommand{\bS}{{\boldsymbol{S}}}
\newcommand{\bk}{{\boldsymbol{k}}}
\newcommand{\br}{{\boldsymbol{r}}}
\newcommand{\tJ}{{\tilde{J}}}
\newcommand{\tK}{{\tilde{K}}}
\newcommand{\tmc}{{\text{mc}}}
\newcommand{\hU}{{\hat{U}}}
\newcommand{\td}{{\text{d}}}

\begin{document}
\title{Magnon and photon blockade in a hybrid antiferromagnet–cavity quantum system}
\author{Vemund Falch} 
\author{Arne Brataas}
\author{Alireza Qaiumzadeh} 

\affiliation{Center for Quantum Spintronics, Department of Physics, Norwegian University of Science and Technology, NO-7491 Trondheim, Norway}
\date{September 17, 2025}

\begin{abstract}
We investigate both magnon and photon blockade for an antiferromagnetic insulator coupled to a linearly polarized cavity mode. We focus on the cross-Kerr nonlinearity between the two magnon modes, which can be large in antiferromagnets with a weak easy-axis magnetic anisotropy. By numerically solving the Lindblad master equations, we demonstrate that the resulting bright and dark modes, i.e., system eigenmodes that couple strongly and weakly to photons, respectively, exhibit distinct behaviors. The bright mode exhibits both magnon and photon blockade due to a weak effective nonlinearity, while the dark mode only exhibits magnon blockade for a detuned cavity photon. The blockade efficiency can further be optimized by appropriately tuning the competing interactions in the system. In addition, we show that applying a DC magnetic field, which lifts the degeneracy of antiferromagnetic chiral magnon eigenmodes, destroys the dark mode and leads to an unconventional photon blockade. These findings provide a pathway for generating single magnon and photon states useful for quantum information technology based on the underlying large squeezing of antiferromagnetic magnons.
\end{abstract}

\maketitle

\section{\label{sec:Introduction}Introduction}
Magnonics, the study of quantized spin waves in magnets, has seen great interest for use in information technology due to the low intrinsic damping and easy interoperability of magnetic insulators with other systems \cite{MagnonSpintronics,MagnonSwitching,SpinInsulatronicsReview}. Cavity magnonics in particular, based on the strong interaction between magnons and cavity photons \cite{MagCavCoupl1,MagCavCoupl2,CavMagnonics}, have attracted significant attention for investigating quantum phenomena, such as squeezed, entangled, and cat states \cite{3CrossKerrEntanglement,MagnonArrayEntanglement,SqueezedStates,MagnonCatState}, as well as single magnon and photon states \cite{SingleMagnonExperiment,HigherOrderBlockades,MagnetoOpticalPB_3BothKerr}, with potential applications in quantum information \cite{QuantumInform1}, quantum sensing \cite{QuantumSensing1, QuantumSensing2, QuantumSensing3}, and quantum transduction \cite{QuantumTransducing, CavityMagnomechanicsTransduction}.

In conventional magnonics, insulating ferromagnets, most prominently yttrium iron garnet (YIG), have been the material of choice due to their exceptionally low magnetic damping and long magnon lifetimes. However, antiferromagnets (AFMs) offer additional advantages due to their THz frequencies, lack of parasitic stray fields, and two chiral magnon modes with opposite handedness, at the cost of being more difficult to control and detect \cite{AFRev1,AFRev2,AFDetection,AFM_Handedness}.  While interactions between antiferromagnetic magnons and cavity photons are normally weakened by antiferromagnetic spin fluctuations \cite{BoventerAFCav,NonLocalFMAFCoupl_Oyvind}, recent advances have demonstrated strong coupling of THz cavities to various systems \cite{USCTwoLevelRes, ThzCavity, ScalariStrongCouple} including optomagnonic AFMs and AFMs in the GHz range \cite{BoventerAFCav,GhZCavity1,MagnetoOptocial3,MagnetoOptocial2,THzAFCavExp,AFM_GHzCouplNew}. A recent proposal predicted that hybrid structures can further enhance the coupling strength \cite{EnhancedAFCavCoupl}, and other nonresonant AFM-cavity coupling schemes have also been investigated \cite{AFCav_Raman,AFCav_Raman_DoubleEl}. This has sparked interest in exploring the physics behind antiferromagnetic cavity magnonics, including magnon dark modes; magnons that either do not couple or couple only weakly to the cavity photon mode \cite{MagnonDarkModes_TwoCavity, AFHam2}, magnon-magnon entanglements \cite{AFLangevin}, enhanced superconductivity \cite{AFSuperconductor}, and enhanced critical fluctuations at the critical point \cite{AFCav_Criticality}.  

The bosonic blockade, analogous to the Coulomb blockade of fermionic charged particles \cite{GorterCoulombBlockade,CoulombBlockade}, arises when the excitation of a single boson suppresses the excitation of additional bosons \cite{PhotonBlockade,Yuan_Antibunching}. This effect can arise from two distinct mechanisms: in a conventional blockade, a nonuniform level spacing suppresses successive resonant excitations \cite{PhotonBlockade, Birnbaum2005, Yuan_Antibunching}, while in an unconventional blockade, destructive quantum interference between different excitation paths inhibits the excitation of a second boson \cite{UncoventionalBlockadeOrig1, UnconventionalBlockadeOrig2}.
Bosonic blockades enable the generation of single bosons, an essential building block for quantum technologies, and have been extensively investigated for photons \cite{SinglePhotonApp1,SinglePhotonApp2,SinglePhotonApp3}. Single photons have been utilized as state-of-the-art quantum computation platforms \cite{SP_QC1,SP_QC2}, and for quantum key distribution \cite{SinglePhotonKeyDistribution}. Single bosons are also important building blocks for generating more complex quantum states, like entangled states or cat states, for applications in quantum metrology \cite{Metrology1,Metrology2} or quantum computing \cite{EntangledStateQC1, EntangledStateQC2,CatStateQC1}.

Recent research has studied the photon blockade in a variety of systems, magnetic \cite{MagnonKerrSinglePhoton1,OldPaper,IsingMagnet} or other hybrid cavity systems \cite{NonreciprocalPhotonBlockade,MagnetoOpticalPB_2BothKerr,OddKerr_2PB}, to enable high-quality generation of single photons. The current state-of-the-art single photon sources are based on semiconductor quantum dots in microcavities \cite{SP_QC1,QDCav}, and the unconventional photon blockade has also been observed \cite{TwoPhotonBlockadeExp,ObservationBlock2,UPB3}. The corresponding magnon blockade has also been investigated in a variety of ferromagnetic cavity systems \cite{MB:YIG_SC_Cav,MagnonCavity_AtomBlockade,MB:YIG_Cav_PA}, with the goal of creating a single magnon source. Recent advances in high sensitivity magnon detection \cite{MagnonSensing,MagnonSensing:arXiv,QuantumSensing1} could also lead to the experimental verification of such states. Higher order blockades, by stabilizing $n$ excited photons or magnons, have also garnered interest \cite{TwoPhotonBlockadeExp,HigherOrderBlockades,MagnonBundle}.

In this paper, we perform a comprehensive study of magnon and photon blockades in a hybrid AFM-cavity system. The system exhibits both magnon self- and cross-Kerr nonlinearities \cite{CrossKerrCoupl_OptoMech,CrossKerrCoupl_OptoMech2,CrossKerr_MagPhotPhonExp}, arising from the spin-spin interactions. Quantum systems with Kerr nonlinearities have been shown to exhibit exotic quantum effects, such as bistability and multistability~\cite{MagnonDriving,CrossKerr_MagPhotPhonExp,TriStable_MemLog}, squeezing~\cite{KerrSqueezing1,KerrSqueezing2}, and entanglement~\cite{Entanglement}, with applications in nonreciprocal transmission~\cite{NonReciprocalTransmission,NonreciprocalPhotonTransmission}, information storage and processing~\cite{TriStable_MemLog,QuantumOpticalGate}, and enhanced coupling~\cite{SpinSpinCoupl,OpticalSpring}. In particular, Kerr nonlinearities have been shown to instigate blockades in other systems, including optomechanical and ferromagnetic cavities with similar effective Hamiltonian structure \cite{MagnetoOpticalPB_3BothKerr,MagnetoOpticalPB_3CrossKerr,3Mode:OptMechanical}. We show that the ratio of self- and cross-Kerr nonlinearity strengths equals the ratio of easy-axis magnetic anisotropy and Heisenberg exchange coupling strengths, influencing the optimal choice of antiferromagnetic materials. By considering the symmetry of the system, we show that the choice of orientation and detuning of the driving field are closely linked. We then relate the observed blockades to the eigenstates of the system and investigate how the blockade efficiency depends on detuning, magnon-photon interaction, and the strength of the Kerr nonlinearities. We further show how competing interactions lead to a nonreciprocal blockade, which can further enhance the blockade efficiency.

The rest of the paper is organized as follows: We present the system model and corresponding bosonic Hamiltonian in Sec.~\ref{sec:model} and discuss how to characterize it using the second-order correlation function in Sec.~\ref{sec:CorrFuncs}. We present numerical results for the correlations of both the cavity photon and magnons, for two different orientations of the driving field in Sec.~\ref{sec:ParPump} and \ref{sec:OrthPump}, and discuss how to optimize the blockade efficiency. Finally, we discuss realistic material parameters and present our conclusion in Sec.~\ref{sec:Conclusion}.

\section{\label{sec:model}System Model}
\begin{figure}[t]
    \centering
    \includegraphics[width=\linewidth]{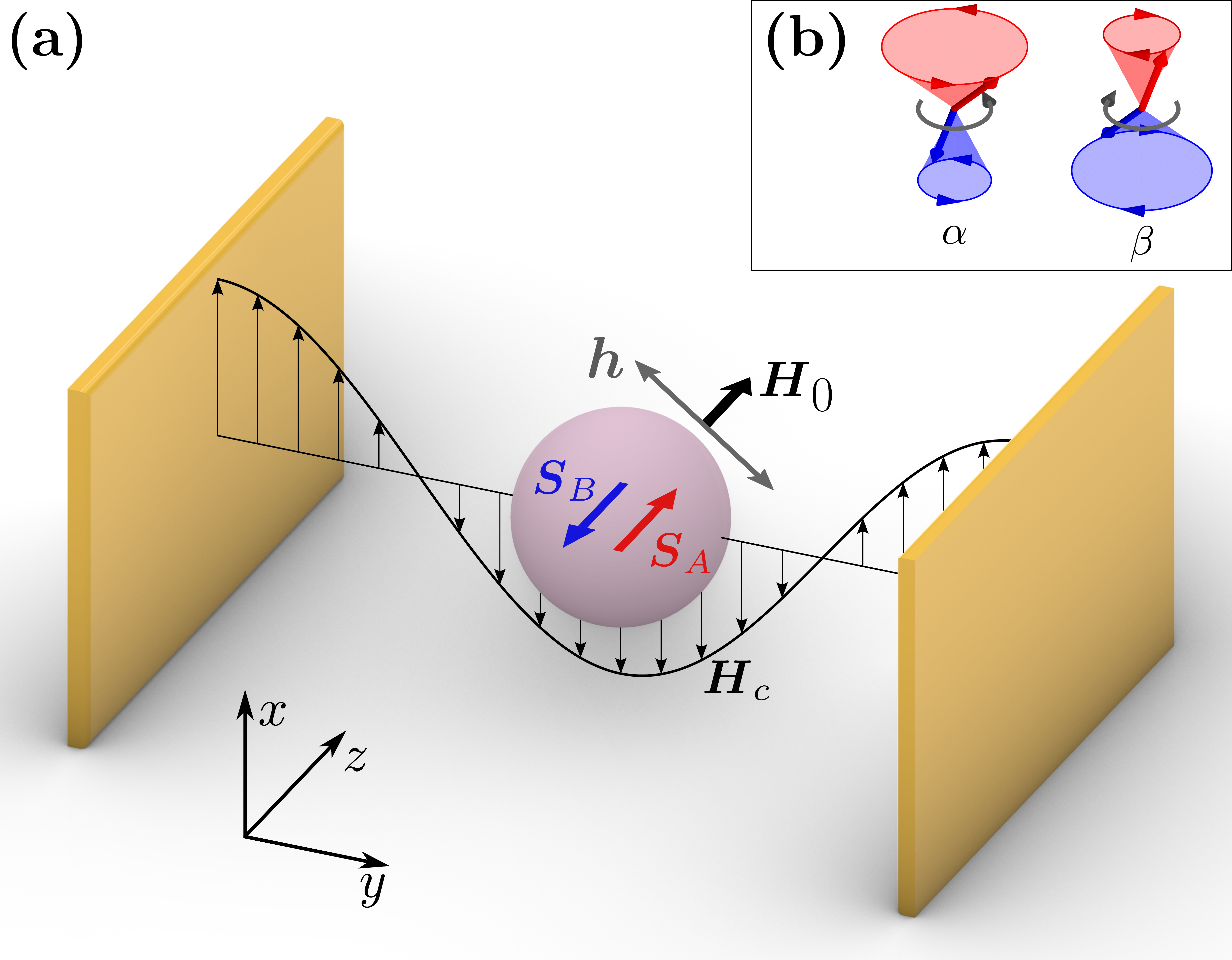}
    \caption{(a) Schematic illustration of the hybrid AFM-cavity system. The cavity supports a linearly polarized electromagnetic mode with magnetic field component $\boldsymbol{H}_c$, which couples to the two localized antiferromagnetic spin sublattices $\bS_A$ and $\bS_B$, via the Zeeman interaction. The AFM is further biased by an external static magnetic field $\bH_0$ parallel to the magnetic ground state and driven by a transverse oscillating external field $\bh$ perpendicular to $\bH_0$. (b) Schematic view of $\alpha$ and $\beta$ magnons modes with opposite handedness \cite{AFMagnons,AFMagnonIll}.}
    \label{fig:SystemOverview}
\end{figure}

In this section, we derive the effective bosonic Hamiltonian of a hybrid AFM-cavity system as illustrated in Fig.~\ref{fig:SystemOverview}(a). The total Hamiltonian of the hybrid system can be separated into $\cH=\cH_m+\cH_c+\cH_\text{int}$ for the AFM magnon modes, cavity modes, and the magnon-cavity interactions, respectively, which we derive in the following.

\subsection{\label{sec:Hamiltonian}AFM Hamiltonian}
We consider a collinear uniaxial easy-axis AFM with two antiparallel magnetic sublattices along the $\hat{\mathbf{z}}$ direction, subjected to an applied DC magnetic field $H_0$ along the magnetic ground state and a transverse AC magnetic driving field $\bh$ \cite{MagnonDriving}, as illustrated in Fig.~\ref{fig:SystemOverview}(a). The spin Hamiltonian of the AFM then reads \cite{AFMagnons,Shiranzaei_2022}
\begin{equation}
	\cH_\text{AFM}=J\sum_{\langle i,j\rangle}\bS_i\cdot\bS_j-K\sum_i(S_i^z)^2-\hbar\gamma\textbf{H}(t)\cdot\sum_i  \bS_i,
\end{equation}
where $J>0$ is the strength of the nearest-neighbor antiferromagnetic Heisenberg exchange interaction, $K>0$ is the single-ion easy-axis magnetic anisotropy energy, $\bS_i$ is the spin vector on the lattice site $i$ with an amplitude $S$, $\hbar$ is the reduced Planck's constant, $\gamma$ is the gyromagnetic ratio and the total external field is $\boldsymbol{H}(t) = H_0 \hat{\boldsymbol{z}} + h_y(t) \hat{\boldsymbol{y}} + h_x(t) \hat{\boldsymbol{x}}$.

The spin Hamiltonian can be rewritten in terms of bosonic modes by a Holstein-Primakoff transformation for the two antiparallel magnetic sublattices $A$ and $B$~\cite{CavMagnonics}
\begin{subequations}
\begin{align}
	&S^z_{i,A}=S-\ha^\dagger_i \ha_i, & S^+_{i,A}\approx\sqrt{2S}\left(1-\frac{\ha_i^\dagger\ha_i}{4S}\right)\ha_i,\\
	&S^z_{i,B}=-S+\hb^\dagger_i \hb_i,&S^+_{i,B}\approx\sqrt{2S}\hb_i^\dagger\left(1-\frac{\hb_i^\dagger\hb_i}{4S}\right),
\end{align}\label{eq:AFHP3}%
\end{subequations}
where $S_i^{+}=(S_i^{-})^\dagger=S_i^x + i S_i^y$ and $\hat{a}_i$ ($\hat{b}_i$) is the bosonic annihilation operator on sublattice $A$ ($B$). After Fourier transforming $\ha_i(\hb_i)=1/\sqrt{N}\sum_\bk e^{-i\bk\cdot\br_i}\ha_\bk(\hb_\bk)$, where $N$ is the number of magnetic unit cells, we can restrict our considerations to the homogeneous modes at $\bk=0$, assuming that the AFM is sufficiently small \cite{Yuan_Antibunching} and the cavity field sufficiently uniform over the AFM \cite{NonUniformCavField}. The quadratic part of the bosonic Hamiltonian then reads
\begin{align}
    \begin{split}  \cH_m^{(2)}=&\varepsilon^a\ha^\dagger\ha+\varepsilon^b\hb^\dagger\hb+\varepsilon^{ab}\big(\ha\hb+\ha^\dagger\hb^\dagger\big),
    \label{eq:FMHamUniform}
    \end{split}
\end{align}
where $\varepsilon^{a(b)}=SJz_1+2SK_z + (-) \hbar \gamma H_0$, $\varepsilon^{ab} = SJz_1$, $z_1$ is the number of nearest neighbors, and $\ha(\hb)=\ha_{\bk=0}(\hb_{\bk=0})$. The quadratic bosonic Hamiltonian \eqref{eq:FMHamUniform} can be diagonalized by the Bogoliubov transformation 
\begin{equation}
\hal=u\ha+v\hb^\dagger,\hspace{1cm} \hbe = u\hb+v\ha^\dagger,\label{eq:AF_Bogo}
\end{equation}
where $u=\cosh(\Theta/2)$ and $v=\sinh(\Theta/2)$ are Bogoliubov coefficients, defined in terms of the Bogoliubov angle $\Theta=\text{arctanh}[(1+2K/Jz_1)^{-1}]$, and $\alpha(\beta)$ denotes the magnon annihilation operator for the right-handed (left-handed) magnon mode with spin-angular momentum $-\hbar$ ($+\hbar$), as illustrated in Fig.~\ref{fig:SystemOverview}(b). The total magnon Hamiltonian at $\bk=0$, up to fourth order in the magnon operators and within the rotating wave approximation \cite{RWA}, becomes
\begin{align}
\begin{split}	\cH_m=\hbar\omega_\alpha\hal^\dagger\hal+\hbar\omega_\beta\hbe^\dagger\hbe-\frac{K}{N}\Big[\big(\hal^\dagger\hal\big)^2+\big(\hbe^\dagger\hbe\big)^2\Big]&\\
	-\frac{Jz_1}{N}\hal^\dagger\hal\hbe^\dagger\hbe+\hbar\Big[\xi_\alpha\hal e^{i\omega_d t}+\xi_\beta\hbe e^{i\omega_d t}+\text{H.c.}\Big]&,
\end{split}\label{eq:AF_Ham}
\end{align}
where $\omega_{\alpha(\beta)}=\omega_0 + (-) \gamma H_0$, with $\omega_0=2S\sqrt{K(Jz_1+K)}/\hbar$, are the bare frequencies of the two magnon modes, and we defined the effective driving strength $\xi_{\alpha(\beta)}=-\gamma\sqrt{SN/8}(1+Jz_1/K)^{-1/4}h_0^{-(+)}$ with $h_0^{\pm}={h}^0_xe^{i\phi_{x}} \pm i{h}^0_ye^{i\phi_{y}}$ assuming a sinusoidal transverse applied AC magnetic field  $h_{x(y)}(t)=h^0_{x(y)}\cos\left(\omega_dt+\phi_{x(y)}\right)$ with driving frequency $\omega_d$ and amplitude $h^0_{x(y)}$. 
In our model AFM, the two magnon modes are degenerate in the absence of the DC magnetic field $H_0=0$.
We have restricted our consideration to the lowest order driving and nonlinear terms as an expansion in the small parameter $(NS)^{-1}$. 

\subsection{Cavity Hamiltonian}
We consider that the antiferromagnetic chiral magnons couple dominantly to a single linearly polarized cavity mode.
The noninteracting Hamiltonian of the single cavity mode is given by 
\begin{equation}
\cH_c=\hbar\omega_c\hc^\dagger\hc,
\end{equation}
where $\hc (\hc^\dagger)$ is the annihilation (creation) bosonic operator of the linearly polarized cavity mode with frequency $\omega_c$. 

\subsection{Interacting AFM-cavity Hamiltonian}
We assume that the AFM is positioned at the maximum of the magnetic field mode of a linearly polarized cavity, so the AFM and cavity photon interact through the Zeeman coupling
\begin{equation}
	\cH_\text{int}=-\hbar\gamma\sum_i\bH_c(\br_i)\cdot\bS_i,
\end{equation}
where $\bH_c(\br_i)=H_c(\br_i)\big(\hc+\hc^\dagger)\hat{\boldsymbol{e}}$ is the magnetic field component of the cavity mode, which is linearly polarized and oriented perpendicular to the magnetic ground state with the polarization unit vector $\hat{\boldsymbol{e}}=\cos\theta\ex+\sin\theta\ey$. The geometry sketched in Fig.~\ref{fig:SystemOverview}(a) thus corresponds to $\theta=0$. Expressing the spins $\bS_i$ in terms of the chiral magnon operators and using the rotating-wave approximation, we find the interaction Hamiltonian
\begin{equation}	\cH_\text{int}=\hbar\hc(g\hal^\dagger+g^*\hbe^\dagger)+\text{H.c.},
\end{equation}
where $g=- \gamma\sqrt{SN/2}(1+Jz_1/K)^{-1/4} H_c^{\rm max}\zeta e^{-i\theta}$ is the interaction coefficient, with $\zeta\in[-1,1]$ a dimensionless measure of the overlap between the cavity magnetic field with the AFM \cite{NonLocalFMAFCoupl_Oyvind}, and $H_c^{\rm max}$ is the maximum amplitude of the cavity mode's magnetic field. If the cavity mode wavelength is much larger than the AFM, we get $\zeta \approx 1$. The coupling $g$ is reduced by a factor of $(1+Jz_1/K)^{-1/4}=u-v$ compared to a ferromagnet-cavity system, due to the two-mode squeezing of the ground state arising from the Bogoliubov transformation in Eq.~\eqref{eq:AF_Bogo} \cite{BoventerAFCav,NonLocalFMAFCoupl_Oyvind}. This reduction is especially pronounced in the limit of strong exchange coupling $Jz_1/K\gg1$.

It is important to note that if the cavity supported circularly polarized modes with magnetic fields perpendicular to the magnetic ground state, the two chiral magnon eigenmodes will couple to separate chiral photon modes with the same handedness. As a result, in such a geometry, there would be no cavity-mediated interaction between the antiferromagnetic magnon modes \cite{NonLocalFMAFCoupl_Oyvind,MagnonDarkModes_TwoCavity}.

\subsection{Time-independent Hamiltonian in rotating frame approximation}
To obtain an effective time-independent Hamiltonian, we go to the frame rotating at the driving frequency $\omega_d$ by making the unitary transformation $\ctH=\hU\cH\hU^\dagger+i\hbar(\td\hU/\td t)\hU^\dagger$ with $\hat{U}=\text{exp}[i\omega_d(\hal^\dagger\hal+\hbe^\dagger\hbe+\hc
^\dagger\hc)]$. The total Hamiltonian $\ctH=\ctH_m+\ctH_c+\ctH_\text{int}$ then consists of a magnon–cavity interaction term and a driving term
\begin{subequations}
\begin{align}
&\ctH=\ctH_\text{mc}+\ctH_d=\big[\ctH^{(2)}_\text{mc}+\ctH^{(4)}_\text{mc}\big]+\ctH_d, \\
&\ctH^{(2)}_\text{mc}=\sum_{\nu=\alpha,\beta,c}\!\!\!\hbar\Delta_\nu\hnu^\dagger\hnu+\hbar\big[\hc(g\hal^\dagger+g^*\hbe^\dagger)+\text{H.c.}\big],\label{eq:HamAF2}\\
	&\ctH^{(4)}_\text{mc}=-\hbar\tilde{K}	\Big[\big(\hal^\dagger\hal\big)^2+\big(\hbe^\dagger\hbe\big)^2\Big]-\hbar\tilde{J}\hal^\dagger\hal\hbe^\dagger\hbe,
	\label{eq:HamAF4}\\	&\ctH_d=\hbar\big(\xi_\alpha\hal+\xi_\beta\hbe\big)+\text{H.c.}\label{eq:HamPump}.
\end{align}
\end{subequations}
Here $\Delta_\nu=\omega_\nu-\omega_d$ is the detuning of the driving frequency for the different modes $\nu=\alpha,\beta,c$. When considering parameter choices where all the bare frequencies are degenerate, we will also make use of the simplified notation $\Delta=\Delta_\nu$ for brevity. Furthermore, $\tilde{K}=K/\hbar N$ and $\tilde{J}=Jz_1/\hbar N$ determine the strengths of the self- and cross-Kerr nonlinearities, respectively, which shift the magnon frequencies proportionally to the number of excited magnons \cite{MagnonKerrEffect}. For typical antiferromagnets $K \ll Jz_1$, and so the self-Kerr nonlinearity $\tK/\omega_{\alpha,\beta} \approx (2NS)^{-1} \sqrt{K/Jz_1}$ is suppressed by a factor $\sqrt{K/Jz_1}$ relative to the bare frequency as compared to the typical self-Kerr nonlinearity in ferromagnets \cite{MagnonKerrSinglePhoton1,Yuan_Antibunching}. On the other hand, the cross-Kerr nonlinearity $\tJ/\omega_{\alpha,\beta} \approx (2NS)^{-1} \sqrt{Jz_1/K}$ that is unique to antiferromagnets due to the strong sublattice exchange coupling is enhanced by a $\sqrt{Jz_1/K}$ factor instead.

The Kerr nonlinearities for a strong easy axis $K\gg Jz_1$, when the $\alpha$ and $\beta$ magnons are mostly localized on sublattices $A$ and $B$, respectively, can be interpreted as the reduction of the exchange field $Jz_1\langle S^z_{B(A)}\rangle$ and magnetic anisotropy field $K\langle S^z_{A(B)}\rangle$ for spin species $A(B)$. This could be realized in some Ising-type van der Waals AFMs \cite{IsingAFM1,IsingAFM2,IsingAFM3}. The situation becomes more complex in the more common weak magnetic anisotropy regime $K\ll Jz_1$, where the ground state is highly squeezed and magnons are delocalized across both sublattices. In this regime, both the self- and cross-Kerr nonlinearities receive contributions from both the Heisenberg exchange and the magnetic anisotropy energies.

\subsection{Renormalized eigenstates}\label{sec:EigenStates}
Thus far, we have described the system in terms of the bare magnon modes, i.e., the \(\alpha\) and \(\beta\) magnons and the cavity mode \(c\), where states \(\ket{n_\alpha n_\beta n_c}\) with a definite number of magnons and cavity photons form a basis for the Fock space of the system. The statistical properties of a weakly driven system are instead best understood in the eigenbasis of the undriven magnon-cavity Hamiltonian $\ctH_\text{mc}=\ctH^{(2)}_\tmc+\ctH^{(4)}_\tmc$, which includes the Kerr nonlinearities. Since $\ctH_\tmc$ commutes with the total number operator $\hn=\hn_\alpha+\hn_\beta+\hn_c$, the eigenstates $\ket{n,i}$ satisfying $\ctH_\text{mc}\ket{n,i}=\tilde{\omega}_{{n,i}}\ket{n,i}$ will be combinations of Fock states with a constant number of excitations $n=n_\alpha+n_\beta+n_c$:
\begin{equation}
    \ket{n,i}=\!\!\!\!\!\sum_{n_\alpha,n_\beta,n_c}\!\!\!\!\!\delta_{n_\alpha+n_\beta+n_c,n}C^{n,i}_{n_\alpha n_\beta n_c}\ket{n_\alpha n_\beta n_c},
\end{equation}
where $i=1,\hdots,(n+1)(n+2)/2$ labels the different states in order of increasing eigenfrequency $\tilde{\omega}_{n,i}$.

\begin{figure}[t]
    \centering
    \includegraphics[width=\linewidth]{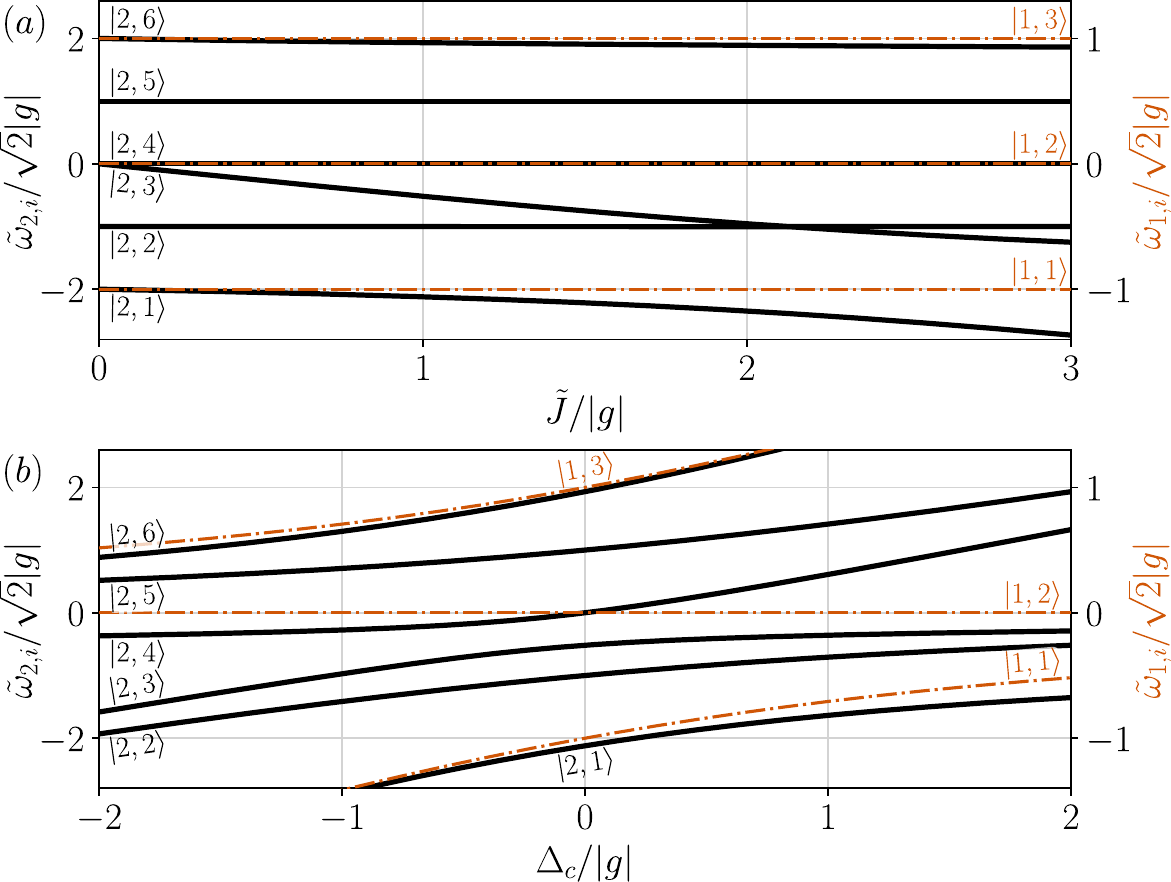}
    \caption{The eigenfrequencies $\tilde{\omega}_{n,i}$ of the eigenstates $\ket{n,i}$ for the coupled magnon-cavity Hamiltonian $\ctH_\tmc$ in the frame rotating at the drive frequency $\omega_d$, as a function of the cross-Kerr nonlinearity $\tilde{J}$ and cavity detuning $\Delta_c$. Only the eigenfrequencies for states with $n=2$ (left axis) and $n=1$ (right axis) excitations are shown. Other parameters are $\Delta_\alpha=\Delta_\beta=\tilde{K}=0$ in both panels, $\Delta_c=0$ in (a) and $\tilde{J}=g$ in (b). }
    \label{fig:EnLvls}
\end{figure}

A conventional blockade occurs when driving the system at a frequency $\omega_d$ where the transition $\ket{0}\to\ket{1,i}$ is resonant for a specific $i$ while all subsequent transitions $\ket{1,i}\to\ket{2,j}$ are off-resonant for all $j$, i.e., $\tilde{\omega}_{i,1}=0$ while $\tilde{\omega}_{2,j}\neq0$. The driving thus efficiently excites a single boson, but further bosonic excitations are suppressed by the frequency mismatch. To study the conditions where a conventional blockade can occur in the hybrid magnon-cavity system, we plot the eigenfrequencies $\tilde{\omega}_{n,i}$ for $n=1,2$ in Fig.~\ref{fig:EnLvls} as a function of $\tilde{J}$ and $\Delta_c$. Note that the scale used for $n=2$ (solid lines) is twice as large as for the $n=1$ (dashed-dotted lines), so that changing $\omega_d$ simply corresponds to shifting the zero point of both $y$ axes in Fig.~\ref{fig:EnLvls} equally. The conditions for a conventional blockade therefore corresponds to parameters where an eigenfrequency $\tilde{\omega}_{1,i}$ does not overlap with any $\tilde{\omega}_{2,j}$'s, and can be realized by appropriately choosing $\omega_d$ to make $\tilde{\omega}_{1,i}=0$ resonant. The strength of the conventional blockade will also correlate with the frequency gap between $\tilde{\omega}_{1,i}$ and the closest $\tilde{\omega}_{2,j}$ states.

From Fig.~\ref{fig:EnLvls}(a) we see that the eigenfrequencies of the single-excitation states are independent of the cross-Kerr nonlinearity $\tilde{J}$, and they can also be shown to be similarly independent of the self-Kerr nonlinearity $\tilde{K}$. They therefore correspond to the quasiparticles of the quadratic Hamiltonian $\ctH^{(2)}_\tmc$, which have been studied previously in Ref. \cite{MagnonDarkModes_TwoCavity}. Importantly, in the degenerate magnon regime $\Delta_\alpha=\Delta_\beta$ the middle state $\ket{1,2}$ is a cavity dark mode, i.e., it has no support in the cavity mode $C_{001}^{1,2}=0$, or equivalently, $\braket{1,2|\hat{n}_c|1,2}=0$. Therefore, $\tilde{\omega}_{1,2}$ is unaffected by detuning the cavity frequency $\Delta_c\neq\Delta_\alpha=\Delta_\beta$, as shown in Fig.~\ref{fig:EnLvls}(b). The dark mode is also antisymmetric with respect to the coupling under sublattice-interchange $\alpha\leftrightarrow\beta$, i.e., $\ket{1,2}=c(g\ket{100}-g^*\ket{010}$) for some constant $c$, while both $\ket{1,1}$ and $\ket{1,3}$ are similarly symmetric in the degenerate magnon regime. This has important implications for the orientation of the driving field to excite the different modes, as we show in Sec.~\ref{sec:NumRes}.

To study the conventional boson blockade, we must therefore investigate the anharmonicity of the eigenfrequencies $\tilde{\omega}_{2,i}$ for the states $\ket{2,i}$ with two excitations. From Fig.~\ref{fig:EnLvls}(a) we see for degenerate bare modes $\Delta_\alpha=\Delta_\beta=\Delta_c$, a finite cross-Kerr nonlinearity causes a frequency gap between both $\ket{1,1}$ and $\ket{2,1}$, and $\ket{1,3}$ and $\ket{2,6}$. Notably, these gaps are not reciprocal, with the former being much larger than the latter. This nonreciprocity can be understood by viewing the cross-Kerr nonlinearity as a perturbation to the quadratic Hamiltonian $\ctH_\text{mc}^{(2)}$, whose eigenstates correspond to states with two excited quasiparticles as discussed in the previous paragraph. A finite nonlinearity then causes both an initial negative frequency shift of, and an effective coupling between, the quasiparticle states with two excitations. The effective coupling in turn causes level repulsion that widens the frequency splitting between the coupled states, increasing (reducing) the initial negative frequency shift for $\ket{2,1}$ ($\ket{2,6}$) and causing the nonreciprocity observed in Fig.~\ref{fig:EnLvls}(a).

The cross-Kerr nonlinearity does not lift the degeneracy of $\ket{1,2}$ and $\ket{2,4}$ in Fig.~\ref{fig:EnLvls}(a), which are always resonant if the magnon frequencies are antisymmetric around $\omega_c$, i.e., $\Delta_\alpha-\Delta_c=-(\Delta_\beta-\Delta_c)$. As such, to open a gap for the dark mode in the degenerate magnon limit $\Delta_\alpha=\Delta_\beta$ requires detuning the cavity $\Delta_c\neq\Delta_\alpha,\Delta_\beta$, as shown in Fig.~\ref{fig:EnLvls}(b). We can again see that the gap between $\ket{1,2}$ and $\ket{2,4}$ is nonreciprocal as a function of $\Delta_c$, which can again be explained by the effective level repulsion in the perturbative picture discussed in the previous paragraph.

\section{Correlation Functions}\label{sec:CorrFuncs}
The hybrid magnon-cavity quantum system can be characterized by the average occupation number of the bosonic modes 
\begin{equation}
\bar{n}_\nu=\langle\hnu^\dagger\hnu\rangle,
\end{equation}
and the equal-time second-order correlation functions
\begin{equation}
	g_{\nu\mu}^{(2)}(0)=g_{\nu\mu}^{(2)}(t,t)=\frac{\big\langle\hnu^\dagger(t)\hmu^\dagger(t)\hmu(t)\hnu(t)\big\rangle}{\big\langle\hnu^\dagger(t)\hnu(t)\big\rangle\big\langle\hmu^\dagger(t)\hmu(t)\big\rangle},
\end{equation}
for both intra-mode ($\nu=\mu$) and inter-mode ($\nu \neq \mu$) correlations between the system’s three bosonic modes $\nu,\mu=\{\alpha,\beta,c\}$, where $\langle\hat{A}\rangle=\text{Tr}\{\hat{\rho}(t)\hat{A}\}$.
The time evolution of the density matrix $\hrho(t)$ can be computed using the Lindblad master equation \cite{CrossKerrCoupl_OptoMech, Yuan_Antibunching}
\begin{align}
	\frac{\partial \hrho(t)}{\partial t}=&-\frac{i}{\hbar}[\ctH,\hrho]\nonumber \\&+\!\!\!\!\sum_{\nu\in \{\alpha,\beta,c\}}\!\!\!\!\kappa_\nu\big[\big(n^\text{th}_\nu+1\big)\cL_\hnu(\rho)+n^\text{th}_\nu	\cL_{\hnu^\dagger}(\rho)\big],\label{eq:LindbladEq}
\end{align}
where the Lindbladian superoperator $\cL_\hnu(\rho)=\hnu\rho\hnu^\dagger-\frac{1}{2}\{\hnu^\dagger\hnu,\hrho\}$ accounts for fluctuations and dissipation in the system, and $\kappa_\nu$ and $n^\text{th}_\nu$ are mode-dependent damping rates and thermal occupation numbers, respectively. For simplicity, we will in the following assume that all the damping rates are equal and introduce $\kappa=\kappa_\alpha=\kappa_\beta=\kappa_c$ as the damping rate of the system. The steady-state density matrix satisfies $\partial_t\hrho=0$. 

In the weak driving or strong blockade regimes, the average occupation number $\bar{n}_\nu$ is approximately the probability of finding a quasiparticle of type $\nu$ in the system. The second-order correlation function describes the conditional probability of finding a second quasiparticle $\mu$ given the presence of a quasiparticle $\nu$. In general, $g^{(2)}(0)>1$ indicates a bunching of the quasiparticles, which are more likely to be excited in pairs, whereas $g^{(2)}(0)<1$ indicates antibunching, where the presence of the first quasiparticle inhibits the excitation of a second quasiparticle  \cite{QuantumOpticsMilburn}. Typically, there is a competition between a large occupation number $\bar{n}_\nu\to 1$ and strong antibunching $g_{\nu\mu}^{(2)}(0)\ll1$ in the blockade regime \cite{HighP1Blockade}.

\section{Numerical Results}\label{sec:NumRes}
In this section, we numerically compute the occupation numbers and correlation functions for the magnon-cavity system described above. Since only the relative angle between the cavity magnetic field $\bH_c$ and driving field $\bh$ matters, we orient the cavity field $\en=\ex$ along the $x$ direction for simplicity, which gives $g = g^* > 0$. We also assume that $\bh$ is linearly polarized and set $\phi_{x}=\phi_y=0$ for simplicity, such that $\xi_\alpha=\xi_\beta^*=\xi$ and the phase of $\xi$ is determined by the angle between $\bh$ and $\bH_c$. We will consider two separate cases: real $\xi=\xi^*$ corresponding to $\bh$ being parallel to $\bH_c$, and imaginary $\xi=-\xi^*$ corresponding to $\bh$ perpendicular to $\bH_c$. We also set the temperature of the bath to $k_BT/\hbar\omega_\nu=0.1$ for all modes $\nu=\alpha,\beta,c$.

All the numerical results are found by solving the Lindblad master equation for the components $\rho_{ikm,jln}=\braket{i_\alpha k_\beta m_c|\hrho|j_\alpha l_\beta n_c}$ of the steady-state density matrix, where we truncate the Fock space to only include components where either all $i,j,k,l,m,n<N_t$ or  $i+j+k+l+m+n<N_s$. To generate the figures, we set \( N_s = 24 \) and \( N_t = 8 \) for the line plots, while for the contour plots, we use \( N_s = 16 \) and \( N_t = 4 \). We have checked that these values of $N_s$ and $N_t$ are sufficiently large to give negligible errors.

\subsection{Driving field parallel to \texorpdfstring{$\bH_c$}{HC}}\label{sec:ParPump}

\begin{figure}[t]
    \centering
    \includegraphics[width=\linewidth]{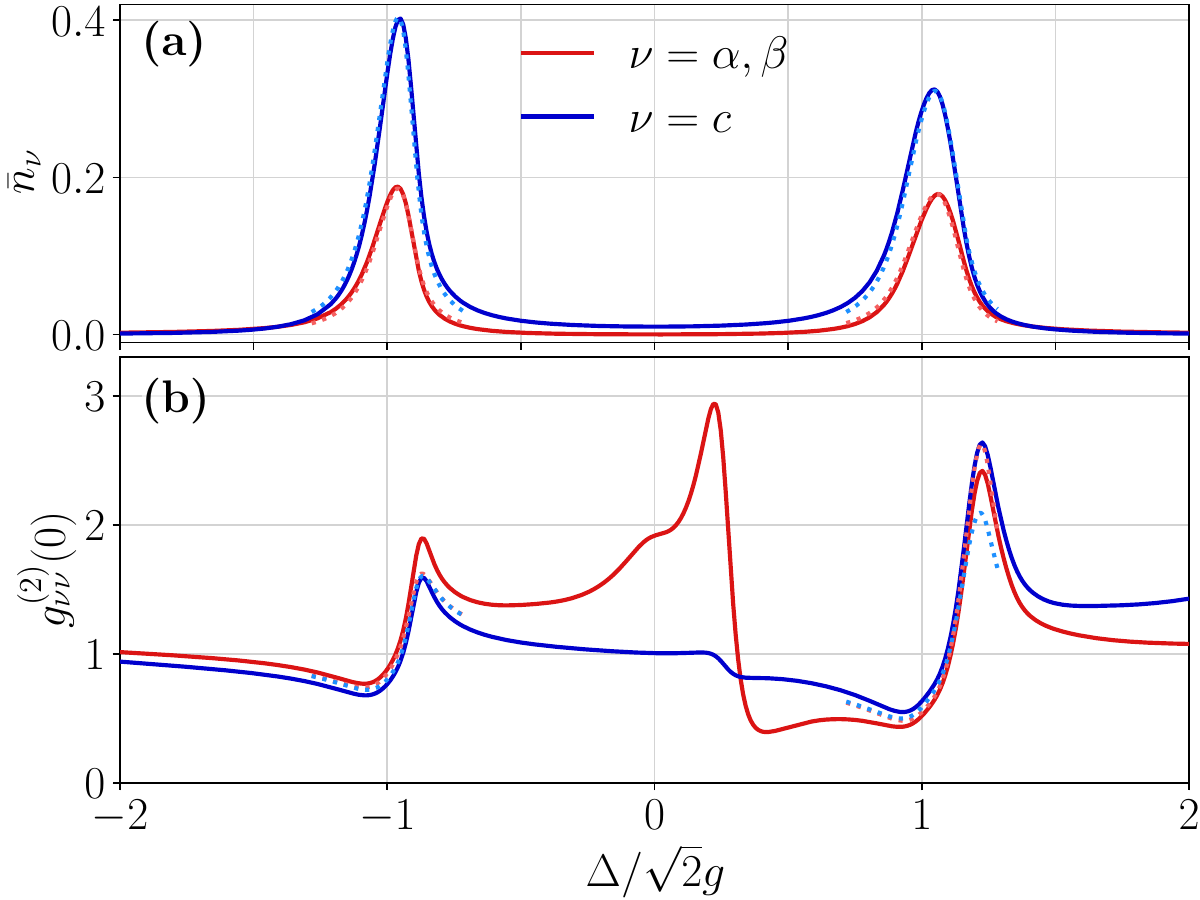}
    \caption{(a) The occupation number $\bar{n}_\nu$ and (b) the intra-mode second-order correlation $g^{(2)}_{\nu\nu}$ for magnons $\nu=\alpha,\beta$ (red lines) and the cavity mode $\nu=c$ (blue lines), as a function of the detuning $\Delta=\Delta_\alpha=\Delta_\beta=\Delta_c$. The dotted lines show the approximate solution when only considering the eigenstates with the highest (lowest) frequency for each $n=0,1,2,3,4$ for $\Delta<0$ ($\Delta>0$). Furthermore, we set the magnon-cavity coupling $g=5\kappa$, cross-Kerr nonlinearity $\tJ=5\kappa$, self-Kerr nonlinearity $\tK=0.05\kappa$, and driving strength $\xi=0.5\kappa$.}
    \label{fig:Delta_Sym}
\end{figure}

We begin by considering the case where the external driving field $\bh(t)$ is along the $x$ axis, i.e., parallel to the cavity magnetic field $\bH_c$, such that $\xi=\xi^*$ is a real parameter. In addition, we set the DC magnetic field to zero $\bH_0=0$, and hence the two magnon modes are degenerate. Figure~\ref{fig:Delta_Sym} shows the resulting occupation numbers $\bar{n}_\nu$ and intra-mode second-order correlations $g^{(2)}_{\nu\nu}(0)$ for all bosonic modes $\nu=\alpha,\beta,c$ as a function of the detuning of the driving frequency $\Delta=\Delta_\alpha=\Delta_\beta=\Delta_c$ when all the bare frequencies are degenerate. The occupation $\bar{n}_\nu$ in Fig.~\ref{fig:Delta_Sym}(a) is large when resonantly driving into the lower $\ket{1,1}$ and upper $\ket{1,3}$ eigenstates for $\Delta\sim\sqrt{2}g$ and $\Delta\sim-\sqrt{2}g$, respectively. However, as the dark mode $\ket{1,2}$ is antisymmetric under interchange of the magnon modes for $g>0$ while the driving is symmetric, the dark mode is not excited even though it is resonant at $\Delta=0$.

At resonance $\Delta=\pm\sqrt{2}g$, we also observe intra-mode antibunching $g_{\nu\nu}^{(2)}(0)<1$ for both the magnon and cavity modes. To study the origins of this blockade we have also solved the Lindblad master equation \eqref{eq:LindbladEq} for a reduced density matrix including only the most resonant eigenstate for each excitation number $n\leq4$, which is shown by the dotted lines in Fig.~\ref{fig:Delta_Sym} in good agreement with the full numerical results. This is therefore a conventional blockade, caused by the anharmonicity from the cross-Kerr nonlinearity shifting the eigenstates with multiple excitations ($n\geq 2$) away from resonance, as seen from the eigenspectrum for $n=2$ in Fig.~\ref{fig:EnLvls}(a). For the same reason, when detuning $\Delta$ above resonance, we see that the modes are bunched $g^{(2)}_{\nu\nu}(0)>1$ as the higher-order transition $n=1\to n=2$ becomes resonant, while $n=0\to n=1$ becomes off-resonant. However, the bunching and antibunching effects are weak, and for $g\gg\tJ$ in particular, the anharmonicity of $\ket{2,1}$ and $\ket{2,6}$ can be shown to only be equivalent to that of an effective nonlinearity $\tilde{J}/16$ in a single-mode system.

\begin{figure}[t]
    \centering
    \includegraphics[width=\linewidth]{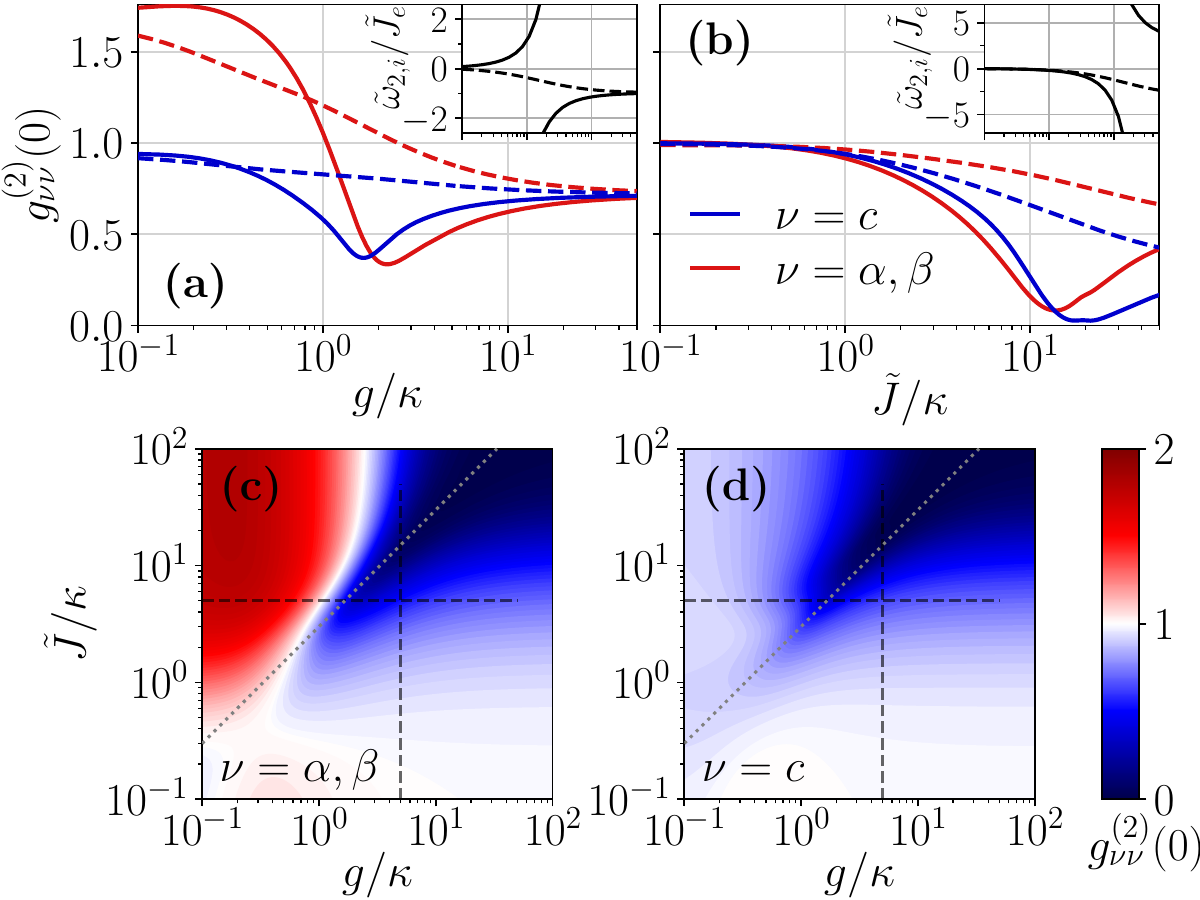}
    \caption{The intra-mode second-order correlation $g^{(2)}_{\nu\nu}$ for both magnons $\nu=\alpha,\beta$ and the cavity mode $\nu=c$ as a function of (a) the magnon-cavity coupling $g$ and (b) cross-Kerr nonlinearity $\tJ$, for detuning $\Delta=\sqrt{2}g$ (solid lines) and $\Delta=-\sqrt{2}g$ (dashed lines). The insets show the eigenfrequencies $\tilde{\omega}_{2,i}$ of the Hamiltonian $\ctH_\tmc$ with the same $x$ axis as the main figures, zoomed in such that only $\tilde{\omega}_{2,1}$ and $\tilde{\omega}_{2,2}$ for $\Delta=\sqrt{2}g$ (solid lines) and $\tilde{\omega}_{2,6}$ for $\Delta=-\sqrt{2}g$ (dashed lines) are visible. The frequencies have been normalized to the effective nonlinearity $\tJ_e=\tJ/8=5\kappa/8$ in (a), with the same constant value also used in (b) for clarity.
    (c) Two-dimensional plot of $g^{(2)}_{\alpha\alpha}=g^{(2)}_{\beta\beta}$ and (d) $g^{(2)}_{cc}$, as a function of both $g$ and $\tJ$ at $\Delta=\sqrt{2}g$.
    The black dashed lines in (c) and (d) correspond to the cuts in (a) and (b), and the gray dotted lines at $\tJ=3g$ are included as a guide to the eye. Other parameters are the same as in Fig.~\ref{fig:Delta_Sym}.}
    \label{fig:OtherPan_Sym}
\end{figure}

We observe in Fig.~\ref{fig:Delta_Sym} that the behavior is not reciprocal for the two detunings $\Delta=\pm \sqrt{2}g$. To further investigate this asymmetry, Figs.~\ref{fig:OtherPan_Sym}(a) and \ref{fig:OtherPan_Sym}(b) show the dependence of the intra-mode correlation functions $g_{\nu\nu}^{(2)}(0)$ on the magnon-photon coupling $g$ and cross-Kerr nonlinearity $\tJ$, respectively, for both $\Delta=\sqrt{2}g$ (solid lines) and $\Delta=-\sqrt{2}g$ (dashed lines). While they behave similarly for both $g\gg\tJ$ and $g\ll \tJ$, for $\Delta=\sqrt{2}g$ the correlation functions $g^{(2)}_{\nu\nu}(0)$ have a pronounced dip at intermediate $g\lesssim \tJ$ that is absent for $\Delta=-\sqrt{2}g$. The nonreciprocal correlations are a consequence of the nonreciprocal eigenfrequencies for $n=2$, which are shown in the insets of Figs.~\ref{fig:OtherPan_Sym}(a) and \ref{fig:OtherPan_Sym}(b) for both detunings, clearly demonstrating that the anharmonicity is greater for $\Delta=\sqrt{2}g$ than for $\Delta=-\sqrt{2}g$. As discussed in Sec.~\ref{sec:EigenStates}, this nonreciprocity is caused by an effective level repulsion caused by the cross-Kerr nonlinearity, increasing (decreasing) the anharmonicity for the state $\ket{2,1}$ ($\ket{2,6}$) relevant at $\Delta=\sqrt{2}g$ ($\Delta=-\sqrt{2}g$). It should be noted that this nonreciprocity also increases (decreases) the weight of the bare state $\ket{110}$ at $\Delta=\sqrt{2}g$ ($\Delta=-\sqrt{2}g$), to such a degree that for $\Delta=\sqrt{2}g$ the magnon inter-mode correlations $1<g_{\alpha\beta}^{(2)}<2$ are weakly bunched in the region of maximal intra-mode blockade.

In Figs.~\ref{fig:OtherPan_Sym}(c) and \ref{fig:OtherPan_Sym}(d) we make a contour plot of the $g$ and $\tJ$ dependence of $g^{(2)}_{\nu\nu}(0)$ for the magnons [Fig.~\ref{fig:OtherPan_Sym}(c)] and cavity mode [Fig.~\ref{fig:OtherPan_Sym}(d)] for $\Delta=\sqrt{2}g$. Consistent with the above analysis, we observe that there is an optimal ratio $\tJ\approx 3g$ that gives the strongest blockade, although interference effects shift the optimal ratio slightly for the magnon modes.

\begin{figure}[t]
    \centering
    \includegraphics[width=\linewidth]{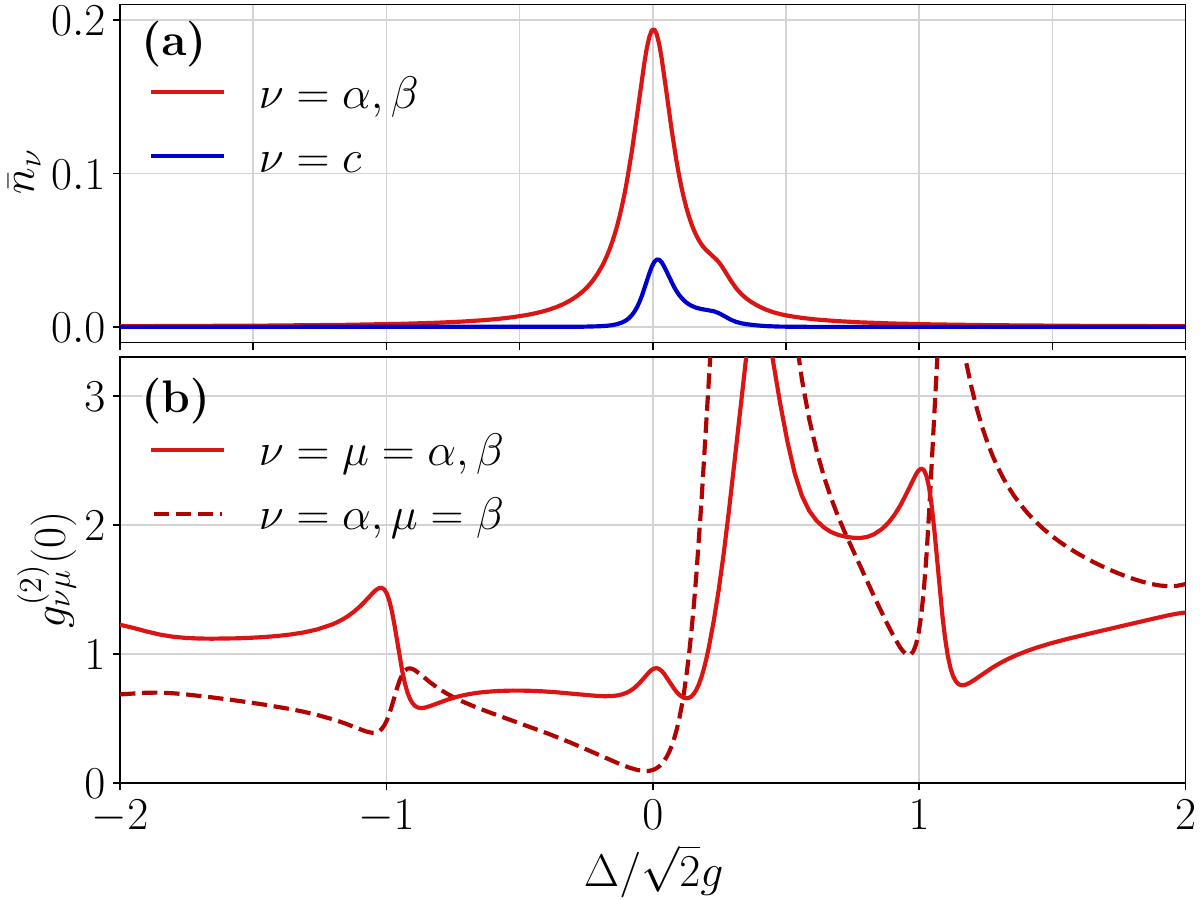}
    \caption{(a) The occupation number $\bar{n}_\nu$ for magnons $\nu=\alpha,\beta$ (red lines) and the cavity mode $\nu=c$ (blue lines) and (b) the magnonic second-order correlation $g^{(2)}_{\nu\mu}$, both intra-mode $\nu=\mu=\alpha,\beta$ (solid line) and inter-mode $\nu=\alpha$, $\mu=\beta$ (dashed line), as a function of the detuning $\Delta=\Delta_\alpha=\Delta_\beta=\Delta_c$. Furthermore, we set the magnon-cavity coupling $g=5\kappa$, cross-Kerr nonlinearity $\tJ=5\kappa$, self-Kerr nonlinearity $\tK=0.05\kappa$, and driving strength $\xi=i 0.3\kappa$.}
    \label{fig:Delta_Asym}
\end{figure}

\subsection{Driving field perpendicular to \texorpdfstring{$\bH_c$}{HC}}\label{sec:OrthPump}
We next consider orienting the external driving field $\boldsymbol{h}$ along the $y$ axis, i.e., perpendicular to the cavity magnetic field such that $\xi=-\xi^*$.
Figure~\ref{fig:Delta_Asym} shows the occupation $\bar{n}_\nu$ and correlations $g^{(2)}(0)$ as a function of the detuning of the driving frequency $\Delta=\Delta_\alpha=\Delta_\beta=\Delta_c$ for degenerate bare modes. We see that the occupation numbers $\bar{n}_\nu$ are resonant around $\Delta=0$ but vanishingly small at $\Delta=\pm\sqrt{2}g$, again explained by the symmetry of the eigenstates. From Fig.~\ref{fig:EnLvls}(a) we see that the behavior at $\Delta=0$ is dominated by the resonant eigenstates $\ket{1,2}$ and $\ket{2,4}$. Interestingly, even though $\ket{1,2}$ corresponds to a cavity dark mode \cite{MagnonDarkModes_TwoCavity}, the cavity mode also appears resonant at $\Delta=0$. The cavity population instead originates from $\ket{2,4}$, which can be expressed as
\begin{equation}
 	\ket{2,4}=\frac{1}{\sqrt{3}}\big(\ket{200}+\ket{020}-\ket{002}\big).
\end{equation}
As such, the probability of finding just a single cavity photon is suppressed, leading to strong bunching $g^{(2)}_{cc}(0)\gtrsim 10$ for the cavity mode, too large to be visible in Fig.~\ref{fig:Delta_Asym}(b) but still finite due to the excitation of nonresonant states and damping-induced quantum jumps populating $\ket{001}$. Furthermore, as $\ket{2,4}$ is at resonance at $\Delta=0$, there is no conventional blockade to give rise to magnon antibunching, and the weak observed antibunching $g^{(2)}_{\alpha\alpha}(0)=g^{(2)}_{\beta\beta}(0)\lesssim 1$ is due to $\ket{2,4}$ being less weighted toward the magnon modes than $\ket{1,2}$. However, as $\ket{2,4}$ has no support in the cross-magnon state $\ket{110}$, the cross correlations $g^{(2)}_{\alpha\beta}(0)\ll 1$ almost vanish.

\begin{figure}[t]
    \centering
    \includegraphics[width=\linewidth]{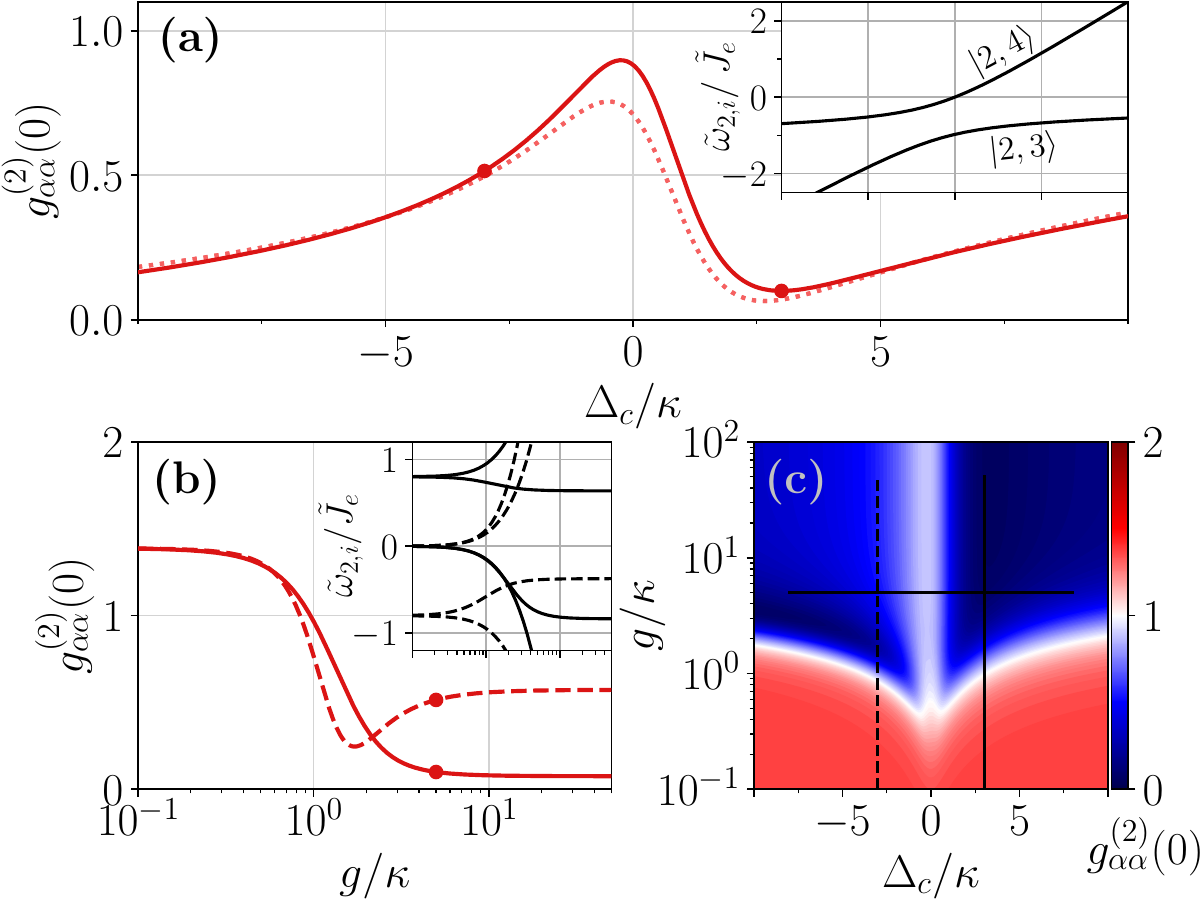}
    \caption{The magnonic second-order correlation function $g^{(2)}_{\alpha\alpha}=g^{(2)}_{\beta\beta}$ as a function of the (a) cavity-detuning $\Delta_c$ at $g=5\kappa$ with an approximate solution only considering the eigenstates $\ket{0},\ket{1,2},\ket{2,3},\ket{2,4}$, (b) magnon-cavity coupling $g$ at $\Delta_c=3\kappa$ (solid lines) and $\Delta_c=-3\kappa$ (dashed lines), and (c) both $\Delta_c$ and $g$. The insets in (a) and (b) show the eigenfrequencies $\tilde{\omega}_{2,i}$ of the Hamiltonian $\ctH_\tmc$ for the eigenstates $\ket{2,i}$ on the same $x$ axes as the main figures, where $\tilde{J}_e=3\tJ/4$. The red dots in (a) and (b) show corresponding equal values of $g$ and $\Delta_c$. The black lines in (c) corresponds to the cuts in (a) and (b).   
     Other parameters are the same as in Fig.~\ref{fig:Delta_Asym}.}
    \label{fig:Other_Wc_Asym}
\end{figure}

As shown in Fig.~\ref{fig:EnLvls}(b), one way to restore the conventional blockade is to only detune the cavity mode $\Delta_c\neq0$ while keeping the bare magnons resonant $\Delta_\alpha=\Delta_\beta=0$. The resulting second-order correlation function is shown in Fig.~\ref{fig:Other_Wc_Asym}(a). We see that a finite detuning increases the magnon blockade by making $\ket{2,4}$ off-resonant, as shown in the inset. We have again verified that this is a conventional blockade by solving the Lindblad equation with a reduced density matrix using only the two eigenstates $\ket{2,3}$ and $\ket{2,4}$ for $n=2$ as well as $\ket{0}$ and $\ket{1,2}$, shown by the dotted line in Fig.~\ref{fig:Other_Wc_Asym}(a) in qualitative agreement with the full result. This blockade is also not reciprocal for $\pm\Delta_c$, which as discussed in Sec.~\ref{sec:EigenStates} is due to the effective level repulsion between $\ket{2,3}$ and $\ket{2,4}$, which increase (decrease) the effective anharmonicity of $\ket{2,4}$ for $\Delta_c>0$ ($\Delta_c<0$) and in turn increase (decrease) the blockade efficiency. As before, it can be shown that this leads to a slight increase in inter-mode correlations $g_{\alpha\beta}^{(2)}\leq 0.3$ as $C_{110}^{2,4}$ becomes nonzero for $\Delta_c\neq 0$.

We further investigate the effect of changing the magnon-photon coupling strength $g$ in Fig.~\ref{fig:Other_Wc_Asym}(b). Similarly to the parallel driving case in Fig.~\ref{fig:OtherPan_Sym}(a), a finite blockade only happens in the strong-coupling regime $g>\kappa$, as in the weak-coupling regime $g<\kappa$ the noise destroys the effective coupling and the system is no longer well described by its eigenstates. We also observe that for small coupling $\kappa\lesssim g<\tJ $, $\Delta_c<0$ actually gives a stronger antibunching due to interference effects similarly to an unconventional boson blockade \cite{UnconventionalBlockadeOrig2}. For large $g\gg \kappa$ this reverses, and we recover the asymmetry of Fig.~\ref{fig:Other_Wc_Asym}(a). In Fig.~\ref{fig:Other_Wc_Asym}(c) we show the full dependence on both $\Delta_c$ and $g$, demonstrating that the above observations also holds for a larger set of parameters.

\begin{figure}[t]
    \centering
    \includegraphics[width=\linewidth]{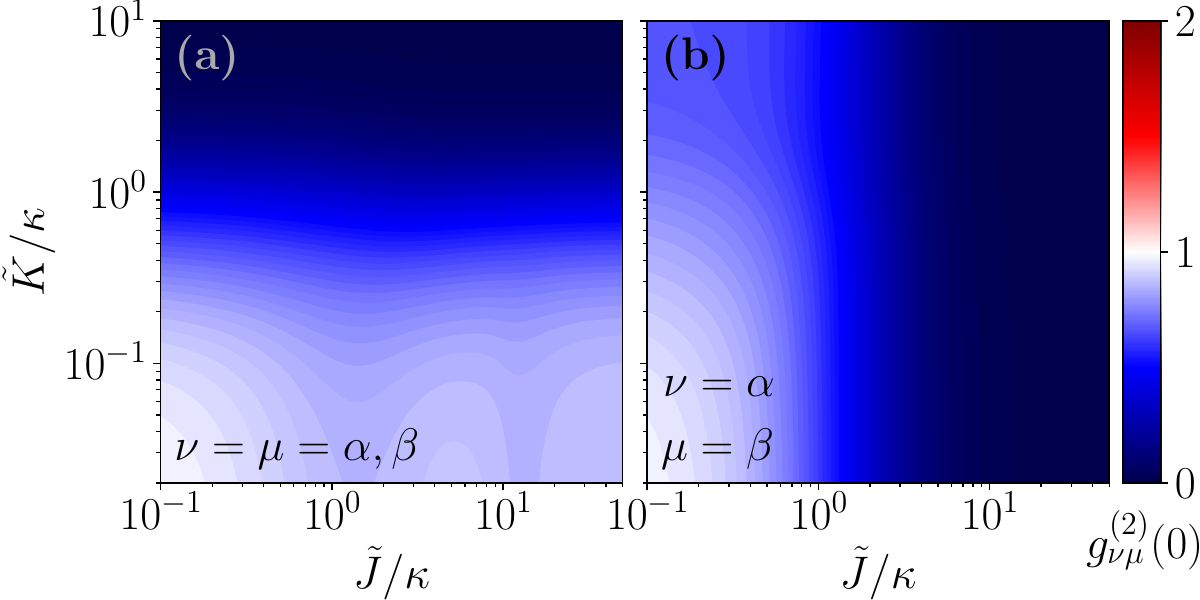}
    \caption{The (a) intra-mode $g^{(2)}_{\alpha\alpha}=g^{(2)}_{\beta\beta}$  and (b) inter-mode $g^{(2)}_{\alpha\beta}$ magnonic second-order correlation functions as a function of the self-Kerr nonlinearity $\tK$ and cross-Kerr nonlinearity $\tJ$ at zero detuning $\Delta_\alpha=\Delta_\beta=\Delta_c=0$. Other parameters are the same as in Fig.~\ref{fig:Delta_Asym}.}
    \label{fig:Other_K}
\end{figure}

To explore the effect of the easy-axis magnetic anisotropy, we plot the dependence of the second-order magnon correlations $g^{(2)}_{\nu\mu}(0)$ on both the self-Kerr $\tK$ and cross-Kerr $\tJ$ nonlinearities in Figs.~\ref{fig:Other_K}(a) and \ref{fig:Other_K}(b), when all bare modes $\Delta_\alpha=\Delta_\beta=\Delta_c=0$ are degenerate and at resonance with the driving. A larger intra-mode nonlinearity $\tK$ lowers the intra-mode correlations $g^{(2)}_{\alpha\alpha}(0)$ and $g^{(2)}_{\beta\beta}(0)$ [Fig.~\ref{fig:Other_K}(a)], while increasing the inter-mode nonlinearity $\tJ$ similarly lowers the inter-mode correlations $g^{(2)}_{\alpha\beta}(0)$ [Fig.~\ref{fig:Other_K}(b)]. It follows that to realize a conventional blockade where $\ket{2,4}$ is anharmonic and not at resonance with $\ket{1,2}$ requires both $\tJ,\tK>\kappa$, otherwise the observed weak inter-mode or intra-mode antibunching stems from shifting the weights of the bare modes in $\ket{2,4}$.

\begin{figure}[t]
    \centering
    \includegraphics[width=\linewidth]{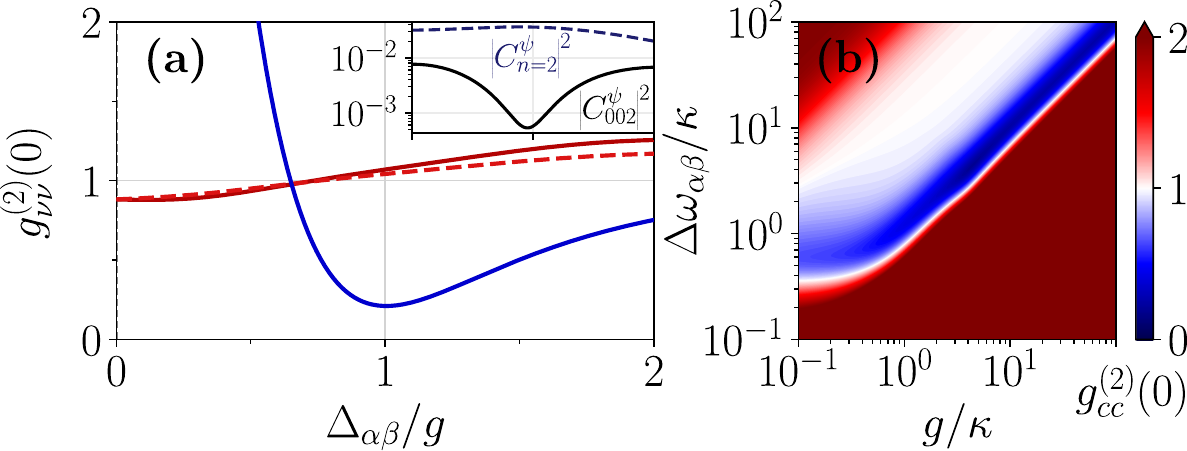}
    \caption{(a) The second-order correlation functions $g^{(2)}_{\nu\nu}(0)$ for both magnons (red lines; $\nu=\alpha$ solid and $\nu=\beta$ dashed) and the cavity $\nu=c$ (blue line) as a function of magnon splitting $\Delta_\alpha=\Delta_{\alpha\beta}$ and $\Delta_\beta=-\Delta_{\alpha\beta}$ at $\Delta_c=0$. (b) $g^{(2)}_{cc}(0)$ as a function of both magnon-cavity coupling $g$ and $\Delta_{\alpha\beta}$. The inset in (a) shows the probabilities of finding two cavity photons $\lvert C_{002}^{\psi}\rvert^2$ (solid line) and two excitations $\lvert C_{n=2}^{\psi}\rvert^2$ (dashed line) in the dominant pure state $\hat{\rho}\ket{\psi}=\rho_{\psi}\ket{\psi}$ of the system, with the same $x$ axis as the main figure. Other parameters are the same as in Fig.~\ref{fig:Delta_Asym}. }
    \label{fig:Other_Wab}
\end{figure}

Lastly, Fig.~\ref{fig:Other_Wab}(a) shows the intra-mode correlations $g_{\nu\nu}^{(2)}(0)$ as a function of the frequency splitting of the magnon modes $\Delta_{\alpha\beta}=\Delta_\alpha-\Delta_\beta$ for antisymmetric magnon detuning $\Delta_\alpha=-\Delta_\beta$ and $\Delta_c=0$. The magnon splitting can be controlled by an externally applied DC magnetic field as shown in Eq.~\eqref{eq:AF_Ham}, but would also be present in thin films due to dipolar interactions \cite{MagnonDarkModes_TwoCavity} or in canted AFMs \cite{BoventerAFCav}. As argued in Sec.~\ref{sec:EigenStates}, for this parameter choice we find $\tilde{\omega}_{1,2}=\tilde{\omega}_{2,4}$, so the transition $\ket{1,2}\to\ket{2,4}$ is resonant and no conventional blockade should occur. This is consistent with the correlation functions for the magnon modes in Fig.~\ref{fig:Other_Wab}(a). However, around $\Delta_{\alpha\beta} = g$ the cavity mode is antibunched due to an unconventional blockade where $C_{002}^{2,4}=\braket{002|2,4}$ vanishes, reducing the probability of finding two cavity photons. We can verify the unconventional nature of the photon blockade by extracting the dominant eigenstate $\ket{\psi}$ of the density matrix $\hrho$, i.e., $\hat{\rho}\ket{\psi}=\rho_\psi\ket{\psi}$ with the largest eigenvalue $\rho_\psi\gtrsim0.9$. In the inset of Fig.~\ref{fig:Other_Wab}(a) we show both the probability $\lvert C_{002}^{\psi}\rvert^2=\lvert\braket{002|\psi}\rvert^2$ of finding two cavity photons and the total probability $\lvert C_{n=2}^{\psi}\big\rvert^2=\sum_{i=1}^6\big\lvert \braket{2,i|\psi}\big\rvert^2$ of finding two excitations. As expected, the probability of finding two cavity photons has a prominent dip at $\Delta_{\alpha\beta}=g$ just like $g^{(2)}(0)$, while the total probability of finding two excitations stays approximately constant.

As in other unconventional blockades, this results from the destructive interference between different excitation pathways \cite{UnconventionalBlockadeOrig2}. However, most other proposals for unconventional blockades involve driving the system off-resonance, resulting in all the eigenstates $\ket{2,i}$ being excited and interfering destructively. This leads to oscillations in the time-delayed correlation function $g^{(2)}(t,\tau)$ due to the eigenstates $\ket{2,i}$ oscillating out of phase, which makes the unconventional blockade more challenging for possible applications \cite{UnconvPhotBlock}. Here we instead drive the system at resonance such that only a single eigenstate contributes dominantly at each $n$, which should reduce the time-delayed oscillations. However, we lose the benefit of operating at weak nonlinearities $\tilde{J}<\kappa$, as a strong nonlinearity is required to suppress transitions into $\ket{2,3}$. In Fig.~\ref{fig:Other_Wab}(b) we have also plotted $g^{(2)}_{cc}(0)$ against splitting $\Delta_{\alpha\beta}$ and magnon-cavity coupling $g$, which shows that the unconventional blockade occurs at $\Delta_{\alpha\beta}=g$ in the strong coupling regime $g>\kappa$, while the noise destroys the correlations in the weak coupling regime $g<\kappa$.

\section{Discussion and concluding remarks}\label{sec:Conclusion}
Our results demonstrate that the magnon and photon blockade depend strongly on the system parameters. Various uniaxial AFMs with distinct material properties have so far been synthesized, which could serve as a promising test bed for our predictions~\cite{AFRev2}. The prototypical antiferromagnetic transition metal difluorides $\text{MnF}_2$, $\text{FeF}_2$, and $\text{CoF}_2$ have anisotropy-to-exchange ratios, $K/Jz_1$, ranging from $K/Jz_1\sim 0.015$ in $\text{MnF}_2$ to $K/Jz_1\sim 0.3$ in $\text{FeF}_2$ \cite{CamleyMnF2_2,AF_Params}, while even smaller ratios have been measured in more exotic antiferromagnetics such as $K/(Jz_1)\sim3\times10^{-4}$ for $\text{Cr}_2\text{O}_3$ \cite{CorondumCr2O3_WeakEasyAxis} or $K/(Jz_1)\sim2\times10^{-5}$ for hematite \cite{Hematite1,Hematite2}. The recently discovered van der Waals magnetic materials with layer-dependent magnetic ordering and controllable spin parameters by external strain and electric fields \cite{zhang20242d,Ebrahimian_2023,PhysRevMaterials.4.094004} may also be good candidates to explore different regimes discussed in this paper. Antiferromagnets can also exhibit narrow linewidths $\kappa_{\alpha,\beta}\sim 10^{-3}\omega_{\alpha,\beta}$ \cite{FeF2_LowGilbert,MnF2Linewidth1,MnF2Linewidth2} that are limited by the crystal quality or domain walls \cite{MnF2Linewidth3}, although the linewidth at a weaker easy axis are generally increased by the stronger two-mode squeezing~\cite{BroadenedLinewidth}. With the representative lattice constant $a=0.5\,\text{nm}$, spin $S=2$ \cite{AF_Params,Cr2O3Spin,Fe2O3Spin}, and $\kappa=10^{-3}\omega_0$, we find that to reach the nonlinear regime $\tilde{J}\gtrsim\kappa$ corresponds to a sphere of diameter $d\lesssim 10\,\text{nm}\,\text{\textendash}\,30\,\text{nm}$ or a thin film with diameter $d\lesssim30\,\text{nm}\,\text{\textendash}\,160\,\text{nm}$ for $K/(Jz_1)\sim10^{-2}\,\text{\textendash}\,10^{-5}$.

Our results also demonstrate that the dominant cross-Kerr nonlinearity in AFMs requires an effective magnon-magnon interaction to be mediated by an external mode for a conventional magnon or photon blockade, in contrast to the self-Kerr interaction in ferromagnets \cite{Yuan_Antibunching}. For the magnon-cavity system under consideration, an effective magnon-magnon coupling is only realized in the strong coupling regime $g>\kappa$. As the coupling strength increases for larger AFMs, this is in tension with the small AFMs required for a strong cross-Kerr nonlinearity. For $\text{MnF}_2$, in particular, strong coupling has been shown to require the volume of the AFM to constitute about $1\%$ of the cavity for $\kappa=10^{-3}\omega_0$ \cite{NonLocalFMAFCoupl_Oyvind}. Furthermore, even in materials with smaller anisotropy-to-exchange ratios, the reduction in the coupling strength from the increased two-mode squeezing perfectly compensates for the allowed size increase of the AFM from the enhancement of cross-Kerr nonlinearity, resulting in the same absolute cavity size requirements as for $\text{MnF}_2$. The strong coupling regime is therefore probably out of reach in regular terahertz cavities when the AFM is size constrained by $\tilde{J}\gtrsim\kappa$, but could be reached in deep-subwavelength cavities \cite{DeepSubwavelengthCav} with very clean AFMs $\kappa \lesssim 10^{-4}$. To increase the magnon-cavity coupling strength, a promising alternative is to engineer the cavity to interact unequally with the two magnetic sublattices, as recently explored in Ref.~\cite{EnhancedAFCavCoupl}. The two-mode squeezing can then result in a drastically increased, although slightly anisotropic, magnon-cavity coupling strength. Similar squeezing-enhanced coupling enhancement could also be realized to other quantum modes beside cavity photons \cite{AFGroundState}, which can be used to induce an effective magnon-magnon interaction, and where our results could serve as an important guide. It could also be noted that degenerate parametric amplification has been demonstrated in easy-plane AFMs \cite{AF_EPlane_DPA}, which could serve as an alternative path to realize antiferromagnetic magnon blockade \cite{QuantumOpticsMilburn}.

In summary, we have theoretically investigated magnon and photon blockade in a hybrid AFM-cavity system, based on the strong cross-Kerr nonlinearity between the antiferromagnetic magnon modes. The cavity bright modes exhibit a weak nonreciprocal conventional blockade due to cross-Kerr nonlinearity that can be optimized by tuning the magnon-cavity interaction strength. On the other hand, the cavity dark mode exhibits a stronger magnon blockade when appropriately detuning the cavity photon. We show how the nonreciprocal detuning dependence follows from an effective level repulsion for the states with multiple excitations. We also find an unconventional photon blockade by tuning the magnon mode splitting via an external magnetic field. Our platform for selectively exciting single magnons could be thus useful for emerging AFM-based quantum information technology \cite{MagMagEntanglAF,QuantSensing,AFCav_Raman}. As the second-order photon correlation function can be measured in a Hanbury Brown-Twiss experiment \cite{HBS}, which has already been done for photon blockades in other hybrid quantum cavity systems~\cite{ObservationBlock2,UPB3,TwoPhotonBlockadeExp}, our results could be experimentally verified if a sufficiently strong magnon-cavity coupling can be realized.

\section*{Acknowledgements}
This project has been supported by the Research Council of Norway through its Centres of Excellence funding scheme, Project No. 262633, ``QuSpin''.

\section*{Data availability}
The data that support the findings of this article are not
publicly available upon publication because it is not technically feasible and/or the cost of preparing, depositing, and
hosting the data would be prohibitive within the terms of this
research project. The data are available from the authors upon
reasonable request.

\bibliography{References}

\end{document}